\documentclass[onecolumn,prc,showpacs,amsmath,amssymb,epsfig,10pt]{revtex4}
\usepackage{epsf,epsfig,graphicx}
\begin{document}
\titlepage
\title{Suspicion on  Engrafting HBT From Astronomy to Heavy Ion Collision }
\author{ X. Sun\\
Institute of High Energy Physics, Beijing 100039, China\\
Graduate University of the Chinese Academy of Sciences, Beijing 100039, China\\
Department of Engineering Physics, Tsinghua University,
Beijing 100084, China\\
Center of High Energy Physics, Tsinghua University, Beijing 100084,
China }

 \begin{abstract}
HBT method in astronomy and heavy ion collision is contrasted in
present article.Some differences are found and validity of using HBT
in heavy ion collision is suspected.
\end{abstract}
 \maketitle
\newpage

\section{Introduction}
The method of two-particle intensity interferometry was discovered
in the early 1950's by Hanbury Brown and Twiss (HBT) \cite{HBT}
who applied it to the measurement of the angular diameter of stars
and other astronomical objects.

Then several
authors~\cite{hbt-review,pratt-flow-and-lifetime,rischke-lifetime,rischke-gyulassy-lifetime,bertsch89,shuryak-lifetime}
have proposed HBT studies to probe  source structure in heavy ion
collision.

The spirit of this  method  is reviewed and  explained by Ulrich
Heinz\cite{heinz99}: "two random point sources $a$ and $b$ on a
distant emitter, separated by the distance $R$, emit identical
particles with identical energies $E_p=(m^2+p^2)^{1/2}$ which, after
travelling a distance $L$, are measured by two detectors 1 and 2,
separated by the distance
%
\begin{figure}[ht]
\centerline{\epsfxsize=13cm\epsffile{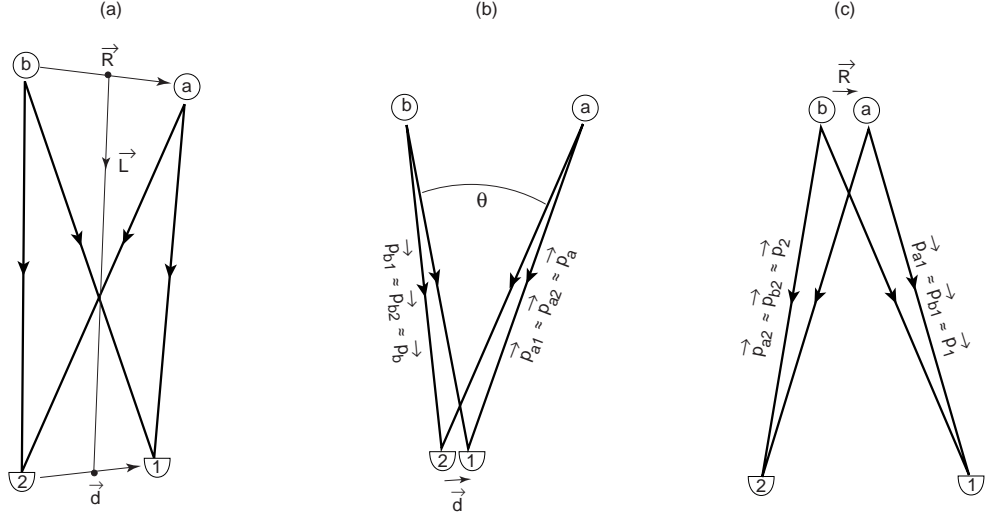}} \caption{Measurement
of the separation $\protect\bbox{R}$ of two
  sources $a$ and $b$ by correlating the intensities in detectors 1
  and 2 at varying distances $\protect\bbox{d}$. {\bf a}: The general
  scheme. {\bf b}: The specific situation in astronomy. {\bf c}: The
  specific situation in particle physics.
\label{F0}}
\end{figure}
%
$\bbox{d}$ (see Figure~\ref{F0}a). $L$ should to be much larger than
$R$ or $d$. The total amplitude measured at detector 1 is then
 \begin{equation}
 \label{B1}
   A_1 = {1\over L}\Bigl( \alpha\, e^{i(pr_{1a}+\phi_a)}
            + \beta\, e^{i(pr_{1b}+\phi_b)} \Bigr),
 \end{equation}
where $\alpha,\beta$ are the amplitudes emitted by points $a$ and
$b$, $\phi_a,\phi_b$ are their random phases, $r_{1a},r_{1b}$ are
their distances to detector 1. For some time ,the two-particle
intensity correlation function is thus given by
 \begin{equation}
 \label{B5}
   C(\bbox{R},\bbox{d}) =
   {\langle I_1 I_2\rangle \over \langle I_1\rangle \langle I_2\rangle}
   ={\langle A_1 A_1^\dagger A_2 A_2^\dagger \rangle \over \langle A_1 A_1^\dagger\rangle \langle A_2 A_2^\dagger\rangle}= 1 + {2 \vert\alpha\vert^2 \vert\beta\vert^2 \over
            (\vert\alpha\vert^2 + \vert\beta\vert^2)^2}
   \cos\Bigl(p(r_{1a}-r_{2a}-r_{1b}+r_{2b})\Bigr).
 \end{equation}
For large $L\gg R,d$, the argument of the second, oscillating term
becomes
 \begin{equation}
 \label{B6}
   r_{1a}-r_{2a}-r_{1b}+r_{2b} \longrightarrow {d\,R\over L}
   \Bigl( \cos(\bbox{d},\bbox{R}) - \cos(\bbox{d},\bbox{L})
          \cos(\bbox{R},\bbox{L}) \Bigr).
 \end{equation}
Note the symmetry of this expression in $d$ and $R$, the
separations of the detectors and of the emitters."this symmetry is
lost in the two practically relevant limits:

\begin{enumerate}

\item
In case of $R\gg d$,the cosine-term in (\ref{B5}) reduces to
$\cos(\bbox{d}{\cdot}(\bbox{p}_a{-}\bbox{p}_b))$, with
$\bbox{p}_{a,b} = p\, \bbox{e}_{a,b}$.

\item
In case of $R\ll d$,Then (see Figure~\ref{F0}c) the cosine-term in
(\ref{B5}) becomes $\cos(\bbox{R}{\cdot}(\bbox{p}_1{-}\bbox{p}_2))$.

\end{enumerate}
So we can get the intensity correlation function
 \begin{equation}
 \label{B7}
   C(\bbox{p}_1-\bbox{p}_2)  =1+
   \int d^3R\, \rho(\bbox{R})\,
   \cos(\bbox{R}\cdot(\bbox{p}_1-\bbox{p}_2))\, .
 \end{equation}

for a continuum of a static sources described by a distribution
$\rho(\bbox{R})$ of their relative distances.

The upper is deduction in theory.Now let's see how to get the
correlation function in experiment.\cite{starpion}"Experimentally,
the two-particle correlation function is obtained from the ratio
$C_2({\bf q}) = A({\bf q})/B({\bf q})$ (normalized to unity at
large ${\bf q}$), where $A({\bf q})$ is the measured two-pion
distribution of pair momentum difference ${\bf q=p_2-p_1}$, and
$B({\bf q})$ is the mixed background
distribution~\cite{event-mixing}, calculated in the same way using
pairs of particles taken from different events."Note ${\bf
p_2,p_1}$ is the primary momentum of particle,i.e. the  momentum
of the particle when the particle is nearest to the source.It
corresponds to the momentum when the particle leaves  the source
in theory.

Really,the variable used in correlation is a Lorentz invariable
$Q_{inv}$.\cite{starpion}"$Q_{inv}=\sqrt{({\bf p}_1-{\bf
p}_2)^2-(E_1-E_2)^2}$".After transforming (\ref{B7}) and $C_2({\bf
q})$  into function of $Q_{inv}$,it is believed that relationship
between experiment and theory is established.

\section{essentials in HBT and Suspicion}
Correlation function in (\ref{B5}) contains five variables
$\bbox{R}$,$\bbox{d}$,$p_1$,$p_2$ and $L$.Correlation function in
(\ref{B7}) contains four variables $\bbox{d}$,$p_1$,$p_2$ and $L$.
The diminishing  of $\bbox{R}$ is caused by the integral over
$\bbox{R}$.The process using HBT equals the process retrievinng
these variables from experiment.

In astronomy, $\bbox{d}$ is  known quantity , $L$ can be got from
other astronomical method.Although measuring $p_1$ and $p_2$ is
difficult,we can cancel them by integral over them that is recording
all particles entering the detector in experiment.So all variables
are got from experiment and HBT succeeds in astronomy.

In heavy ion collision ,in order to retrieve the correlation
function from experiment,these four variables $\bbox{d}$,$p_1$,$p_2$
and $L$ must be fixed.We can get $p_1$ and $p_2$ from detector
easily.The following and last question is to get  $\bbox{d}$ and
$L$,in other words,where the position the particles are detected is.
The thought that we can cancel $\bbox{d}$ and $L$ after the integral
over them in experiment is wrong because we can't place point
detectors throughout  the space and even if we place point detectors
throughout the space in ideal we still can't make this integral in
experiment just because that particle entering the second detector
after entering the first detector has changed its quantum property
by the first detector so (\ref{B1}) becomes invalid. Another opinion
regarding the  first point left in TPC as the position the particle
be detected is also wrong because that the reconstructed track of
particle is classical and the point is on this classical track ,as
well as (\ref{B1}) becomes invalid.

If $\bbox{d}$ and $L$ can't be got from experiment,HBT can't be used
in heavy ion collision.Now let's see what we got from experiment in
\cite{starpion}.The process in experiment is :firstly ,determining
the momentum and position of particle by the points left in
detector;secondly, reversing the particle along the track,by the way
the classical concept of orbit are needed,to the nearest point in
the track to the source and calculating the momentum  at the
point;thirdly,obtaining
 the correlation function $C_2({\bf q}) = A({\bf q})/B({\bf q})$ (normalized to
unity at large ${\bf q}$), where $A({\bf q})$ is the measured
two-pion distribution of pair momentum difference ${\bf q=p_2-p_1}$,
and $B({\bf q})$ is the mixed background
distribution~\cite{event-mixing}, calculated in the same way using
pairs of particles taken from different events. From these process
we can know the momentum used in calculating the correlation
function is the momentum when particle is in the point that is
nearest to the source.This point corresponds to the point in which
the particle emits from the source in theory.

Now we can reply the question  how to get  $\bbox{d}$ and $L$,in
other words,where the position the particles are detected is .In
\cite{starpion},the particles are detected on the surface of the
source and $L$=0 ,$\bbox{d}\sim 0$. So the deduction in
\cite{heinz99} is disabled here.

\section{formula for experiment}
Now let's see what we have done in data analysis in \cite{starpion}.
Define $P(\theta,\varphi,\tau)$ is the emitting probability of the
source where $\theta,\varphi$ is polar angle and azimuth angle
respectively and $\tau$ is the proper time of the source. For an
event with $M$ particles considered,$n_s,_i(q,q+dq)$ is the number
of particles whose momentum difference with the i-th particle
locates $q$ and $q+dq$,where $q$ is ${\bf q=p_2-p_1}$ or $q_{inv}$
in different cases. $n_m,_i(q,q+dq)$ is the number of particles in
other event whose momentum difference with the i-th particle locates
$q$ and $q+dq$, it is like $n_s,_i(q,q+dq)$ expect that
$n_m,_i(q,q+dq)$ is for different events whereas $n_s,_i(q,q+dq)$
for same event.

For a completely random source , such as  an artifical source
produced by computer obeying the distribution
$P(\theta,\varphi,\tau)$ of real data,  it is easy to know

\begin{eqnarray}
 \label{average}
\bar{n}_s,_i(q,q+dq)=\bar{n}_m,_i(q,q+dq)\\
C_2(q)=\frac{\sum_{i=0}^M\bar{n}_s,_i(q,q+dq)}{\sum_{i=0}^M\bar{n}_m,_i(q,q+dq)}=1
\end{eqnarray}
in which  average is over all events. By now,it is clear that the
correlation function has no direct relation with the emitting
probability $P(\theta,\varphi,\tau)$.

In real data,if there are some mechanisms can change
$n_s,_i(q,q+dq)$ to $n_s,_i(q,q+dq)+n_c,_i(q)$,whereas the sum
number of particles and the distribution have not changed so
$\bar{n}_m,_i(q,q+dq)$ have not changed, we can get
\begin{eqnarray}
 \label{mechanism}
C_2(q)=\frac{\sum_{i=0}^M[\bar{n}_s,_i(q,q+dq)+\bar{n}_c,_i(q)]}{\sum_{i=0}^M\bar{n}_m,_i(q,q+dq)}
\end{eqnarray}
$\bar{n}_m,_i(q,q+dq)$ can be got from simulation by computer
after knowing the distribution
$P(\theta,\varphi,\tau)$.$n_c,_i(q)$ is a variable connecting to
physical process on the surface of the source i.e. what happens
when particle is produced from partons. Let $q=\bf
q=p_2-p_1$,(\ref{mechanism}) become the result in \cite{starpion}.
\section{mechanisms produce $n_c,_i(q)$ for small $q$}
As what are interesting is the performance of correlation function
at $q \sim 0$,i.e. the difference of momentums is small,only this
part is discussed here.

 The meaning of $n_c,_i(q)$ is the variance of number of
particles whose momentum $p$ satisfies $p-p_i=q$  from completely
random distribution.If the i-th particle was produced,the
probability of production of particle whose momentum is close to
$p_i$ would increase.In this case $n_c,_i(q)>0$.Now if the i-th
particle was produced,that probability  would decrease.In this
case $n_c,_i(q)<0$.

Let's see some mechanisms making $n_c,_i(q)$ positive or negative.
On the surface of  the source,if a fermion is produced,it will
hold a state of specified momentum  and prevent particle of this
kind produced after it entering this state.
Qualitatively,$n_c,_i(q)$ will become negative. Interaction will
also affect $n_c,_i(q)$.If there an attractive force between a
kind of particles,$n_c,_i(q)$ will become positive,vice
versa,attractive force leading negative $n_c,_i(q)$.

\section{Conclusions}
The most important problem in engrafting HBT from astronomy to
heavy ion collision is the position where particle is detected can
not correspond to corresponding variables in quondam theory.The
momentum used in HBT is that when particle leave the source.At
this point $L$=0 ,$\bbox{d}\sim 0$. So the deduction  is disabled
for experiment.

{\bf Acknowledgments:} We thank  Dr J. Fu for supporting us in many
respects and their constant helps. We thank Prof. J. Li and Z. Zhang
for the earnest supervision. We thank Prof.Shaomin Chen and Yuanning
Gao for the fruitful discussions in this work. The work was
supported in part by the grants NSFC 10447123.


\end{document}